\documentclass[graybox]{svmult}

\usepackage{type1cm}       
                           
\usepackage{makeidx}      
\usepackage{graphicx}       
                            
\usepackage{multicol}        
\usepackage[bottom]{footmisc}
\usepackage{newtxtext}       
\usepackage{newtxmath}       

\usepackage[scaled=0.9]{helvet}

\makeindex

\usepackage[
style=phys,
maxbibnames=10,
sorting=none
]{biblatex}
\DeclareFieldFormat{labelnumberwidth}{{{#1}.\addspace\addspace}}

\begin{document}

\title*{Plasmon Enhanced Spectroscopy and Photocatalysis}
\author{Sajal Kumar Giri and George C. Schatz}
\institute{S. K. Giri \at Department of Chemistry, Northwestern University, 2145 Sheridan Road, Evanston, Illinois 60208, United States
\and G. C. Schatz \at Department of Chemistry, Northwestern University, 2145 Sheridan Road, Evanston, Illinois 60208, United States, \email{g-schatz@northwestern.edu}}

\maketitle

\begin{abstract} 
\
This study examines the Raman scattering and charge transfer properties of molecules adsorbed on the surface of a tetrahedral Au$_{120}$ nanoparticle based on the time-dependent density functional tight-binding (TD-DFTB) method. 
We study Raman scattering (SERS) enhancements for pyridine where the molecule is adsorbed either on the tip (V-complex) or surface (S-complex) of the nanoparticle. 
The scattering intensity is enhanced by a factor of 3-15 due to chemical effects while significantly larger enhancements (in the order of $10^2-10^4$) are observed for plasmon resonance excitation at an energy of 2.5 eV depending on the adsorption site. 
Furthermore, we demonstrate charge transfer between the nanoparticle and a fullerene-based molecule after pulsed excitation of the plasmon resonance which shows how plasmon excitation can lead to negative molecular ion formation. 
All of these results are consistent with earlier studies using either TD-DFT theory or experimental measurements.\\

\end{abstract}

\section{Introduction}
Surface-enhanced Raman scattering (SERS) is a widely used phenomenon among various research communities where the scattering intensity of molecules can be increased by several orders of magnitude when they are adsorbed on the surface of plasmonic nanoparticles.\cite{SERS-VanDuyne-1997,SERS-Creighton-1977,SERS-VanDuyne-2008} 
The enhancement factors are as large as $10^{10}-10^{11}$ and therefore SERS can be studied on the single-molecule level.\cite{SERS-Emory-1997,SERS-Feld-1997,SERS-Schatz-2008,SERS-Kneipp-2008,SERS-CHeng-2019} 
The unexpected enhancement in the Raman signal from pyridine adsorbed on an electrochemically roughened silver electrode was the first evidence of SERS,\cite{SERS-McQuillan-1974} and this was attributed to the extensive surface coverage of molecules on the rough nanostructured surface, but later it was identified as a plasmon-based phenomenon.\cite{SERS-VanDuyne-1997,SERS-Creighton-1977}  
SERS has been successfully applied to several fields including biosensing\cite{Biosensing-VanDuyne-2003,Biosensing-Zhao-2008}, drug analysis,\cite{Drug-Schatz-2009} and electrochemistry\cite{Electrochemical-Zen-2015,Electrochemical-VanDuyne-2015}.  \\

Raman scattering arises because the electric field of the incident electromagnetic radiation induces polarization in irradiated molecules and the component of the charge oscillation at the shifted Raman frequencies leads to emission of scattered light. 
The intensity of the scattered light depends on two components: incident light intensity and the polarizability of the molecule. 
In SERS, the weak Raman intensities for isolated molecules are increased in the presence of plasmonic nanoparticles due to amplification of the oscillating electric field at the molecule that arises from plasmon excitation. 
Thus the treatment of SERS includes the interaction of light with both molecules and particles.  \\ 

Broadly speaking, the two possible mechanisms that describe SERS involve electromagnetic and chemical effects. 
Localized surface plasmon resonance (LSPR) excitation of the conduction electrons in metal nanoparticles of Au, Ag, and Cu with dimensions of 5-200 nanometers generates an enhanced electromagnetic field at the interface of the particle.\cite{LSPR-VanDuyne-2007} 
This enhanced field can modulate several optical processes at the surface including photocurrent,\cite{Photocurrent-Schatz-2009} luminescence,\cite{Exciton-Achermann-2010} and Raman scattering.\cite{SERS-VanDuyne-1997,SERS-Creighton-1977,SERS-VanDuyne-2008} 
Raman scattering scales with the local field as $|E_{loc}(\omega)|^4/|E_0(\omega)|^4$, where the numerator is the localized electromagnetic field and the denominator is the incident field.\cite{SERS-VanDuyne-2008} 
SERS enhancement factors due to electromagnetic effects can reach values up to $10^{9}$.\cite{SERS-Schatz-2008} \\

On the other hand, the chemical enhancement includes the effect of orbital overlap that occurs when molecules adsorb on the surface.
Static charge transfer associated with adsorption modifies the Raman polarizability of the molecule, leading to a contribution to SERS.
Charge transfer (CT) excitations at optical frequency can also contribute to enhancements through a resonance Raman mechanism.\cite{Electronic-Schatz-2008} 
Enhancements due to static chemical effects can be up to a factor of 3-20 as demonstrated theoretically for pyridine adsorbed on the surface of a small tetrahedral Au particle.\cite{TDDFT-Schatz-2006}.  
Resonance CT can lead to larger enhancements, but usually the resonance CT and LSPR frequencies are sufficiently different, or the resonance CT transition is sufficiently broad, that one only has significant SERS enhancement from plasmon excitation and static CT. \\

Atomistic simulations based on quantum mechanics provide insight into the possible SERS mechanisms and enhancements, in addition to providing Raman shifts and adsorbate structures and reactions.
Previous theoretical works have addressed SERS for pyridine adsorbed on the surface of Ag and Au particles with enhancement factors up to $10^2-10^4$ based on the time-dependent density functional theory (TD-DFT).\cite{TDDFT-Schatz-2006,TDDFT-Schatz-2006}    
However, the size of the nanoparticles that can be described with electronic structure theory has been limited to a small number of atoms ($\leq20$) due to the large computational complexities associated with the calculations.
It is therefore a subject of significant importance to explore SERS for larger nanoparticles. Typically this should involve hundreds or even thousands of atoms as quantum size effects lead to noticeable blue shifts of plasmon frequencies with decreasing particle size for particles with less than 100 atoms.
Semi-empirical methods based on INDO have been demonstrated successfully for the efficient simulation of SERS for Ag-pyridine complexes.\cite{HyperRaman-Schatz-1992,INDO-Schatz-2017} 
As an alternative to the INDO calculations, the density functional tight-binding (DFTB) method may also provide efficient solutions for larger complexes.\cite{DFTB-Aradi-2020} 
We explore this possibility in this work. \\

Two recent papers provide details of the DFTB method.\cite{DFTB-Aradi-2020,RTDFTB-Sanchez-2020} 
DFTB is a semi-empirical method which contains many features that are similar to the earlier INDO method, but it is developed from a DFT perspective, and with much greater functionality.
As a consequence, this method provides results that are often comparable to DFT but with significantly lower computational costs and therefore the ground and excited state properties of larger complexes can be simulated efficiently.  
Previous works have demonstrated the success of the DFTB method in various applications including biomolecules,\cite{DFTB-Bio-Wong-2019} metals,\cite{DFTB-Metal-2019,DFTB-Plasmon-Lucas-2020,DFTB-Plasmon-Sala-2022} semiconductors,\cite{DFTB-SemiConductor} plasmonic photocatalysis,\cite{PDC-Sanchez-2022,PDC-Schatz-2023,PDC-Yu-2023}, etc.
However, the accuracy of the prediction is limited by the quality of the parameters that describe the Hamiltonian ($H$) and overlap ($S$) matrices. 
To elevate accuracy and enable applications to a larger portion of the periodic table, new parameters need to be developed and further optimized. \\

The real-time time-dependent density functional tight-binding (RT-TD-DFTB) method allows us to study plasmon-driven charge and energy transfer processes at the metal-molecule interface by simulating laser-induced electron dynamics.\cite{RTDFTB-Sanchez-2020,PDC-Sanchez-2022,PDC-Schatz-2023} 
In these calculations, hot carriers generated through the decay of LSPR states can be transferred to the adsorbed molecules, initiating reaction dynamics on the surface of the metal nanoparticles. 
This includes several processes like molecular bond dissociation and formation,\cite{PDC-Schatz-2023,PDC-Yu-2023} charge transfer (CT),\cite{Hot-Carrier-Erhart-2022} energy transfer,\cite{PDC-Sanchez-2022} spectral shift,\cite{SERS-PCBM-Schatz-2019} etc. 
In the following, we study CT processes between Au$_{120}$ and a fullerene (C$_{60}$) after exciting plasmons at 2.5 eV with an external laser pulse. 
This work was inspired by a recent study from the Van Duyne group, where the fullerene-based molecule [6,6]-phenyl-C$_{61}$-butyric acid methyl ester (PCBM) was found to form negative ions in the presence of gold dimers when irradiated at 532 nm as revealed by SERS measurements. 
Given this, the present study addresses the ability of the DFTB method to study both SERS and plasmon-driven electron transfer processes at the metal-molecule interface. \\

The paper is divided into two parts, in the first part, we explore Raman scattering for  Au$_{120}$-pyridine complexes and the second part contains plasmon-driven CT in a Au$_{120}$-C$_{60}$ complex. 
For Raman scattering, we discuss enhancements due to chemical and plasmon excitation at an energy of 2.5 eV. 
The molecular Raman spectrum obtained with the DFTB method is compared with the B3LYP/TZVP level of DFT theory.
For the CT study, we demonstrate results with RT-TD-DFTB calculations. 
Thus these two separate studies each addresses key benchmarks for using DFTB in future studies where the two calculations are combined.  \\

\section{Computational Details}
All calculations in this work have been carried out with the DFTB+ and ORCA codes. 
To perform time-dependent density functional tight-binding (TD-DFTB) calculations, we use auorg/auorg-1-1 Slater-Koster parameters for gold-molecule complexes downloaded from the dftb.org site. 
To elucidate the accuracy of this method for predicting Raman spectra, we compare DFTB results with conventional results obtained with the B3LYP/TZVP level of DFT theory. 
The self-consistent-charge density functional tight-binding (SCC-DFTB) approach has been used to compute the ground state properties of all the complexes studied.  
To treat the dynamics induced by external driving fields, we use the time-dependent DFTB scheme that is an extension of SCC-DFTB to the time domain.  \\

The Raman scattering activity $S_p$ for a vibrational mode $p$ is calculated from the polarizability derivatives as $S_p=45\alpha_p'^2+7\gamma_p'^2$, where $\alpha_p'$ and $\gamma_p'$ are derivatives of the isotropic and anisotropic polarizabilities with respect to the normal mode $p$. 
The derivatives of the frequency-dependent polarizability are determined from single-point computations for small Cartesian coordinate displacements from the equilibrium geometry along the normal modes for both DFT and DFTB calculautions.
We compute polarizabilities in the frequency domain with a value of the damping parameter $\Gamma=0.15$ eV. 
To obtain a smooth Raman spectrum as a function of Raman shifts, the scattering activity is broadened by a Lorentzian function with a $20$ cm$^{-1}$ width. \\

In real-time TD-DFTB (RT-TD-DFTB) the one-body reduced density matrix is propagated according to the Liouville-von Neumann equation in a nonorthogonal basis\cite{RTDFTB-Sanchez-2020} 
\begin{equation}
    \dot{\rho}=-{\rm i}(S^{-1}H\rho-\rho HS^{-1})
\end{equation}
starting from an initial state defined by Hamiltonian ($H_0$), overlap ($S_0$), and density ($\rho_0$) matrices.
Here $S^{-1}$ represents the inverse of $S$.
The Hamiltonian in the above equation includes a field-matter coupling term $V_A(t)=-\boldsymbol{\mu}\cdot\mathbf{E}(t)$, where $\mathbf{E}(t)$ is the external field and $\boldsymbol{\mu}=\sum_A\Delta q_A \mathbf{R}_A(t)$ is the dipole with atomic coordinate $\mathbf{R}_A$ and charge $\Delta q_A$ for atom A at time $t$.
We consider a time-dependent pulse of the following form to drive the system  
\begin{equation}
    E(t)=E_0{\rm exp}(-(t-t_c)^2/T_l^2)\sin(\omega_lt)
    \label{pulse_eq}
\end{equation}
with field amplitude $E_0$, center of the pulse $t_c$, pulse duration $T_l$, and frequency $\omega_l$.
For calculating spectra, the density matrix evolves in time due to purely electronic dynamics where the nuclei are ``clamped" in space. 
This treatment has been extensively used to study optical processes of material and molecules induced by external time-dependent electric fields.\cite{RTDFTB-Sanchez-2016,RTDFTB-2-Sanchez-2016,RTDFTB-Sanchez-2020}
The Mulliken population is calculated as 
\begin{equation}
    q_A={\rm Tr}_A[\rho S]
\end{equation}
where ${\rm Tr}_A$ denotes trace over orbitals centered on atom A, and the $S$ matrix depends on the coordinates that are parameterized for each pair of orbitals.   \\

We compute all absorption spectra reported here in the time domain.
The density matrix is propagated in time with an external Dirac $\delta$ perturbation of the form $E(t)=E_0\delta(t)\hat{e}$ with field amplitude $E_0$ and direction vector $\hat{e}$. 
The system evolves in time freely and the induced dipole contains information related to the transition energy and couplings between the ground and excited states.
The absorption spectrum is obtained from the imaginary component of the Fourier-transformed dipole moment  
\begin{equation}
    \sigma_{abs}(\omega)=\frac{4\pi\omega}{3c}{\rm Im}\{{\rm Tr}[(\mu(\omega)-\mu_0)/E_0]\}
    \label{abs_spec_eq}
\end{equation} \\
with an initial permanent dipole $\mu_0$, and speed of light $c$. 
This method reproduces linear absorption spectra that are conventionally obtained in the frequency domain in the limit of a weak field intensity, $E_0\rightarrow 0$, that induces linear photoabsorption. 
To broaden the absorption lines in energy, we damp the time-dependent dipoles with a damping function ${\rm exp}(-t/\tau)$ with a decay timescale of $\tau=6$ fs. \\

\begin{figure}[h]
    \centering
    \includegraphics[width = 0.8 \textwidth]{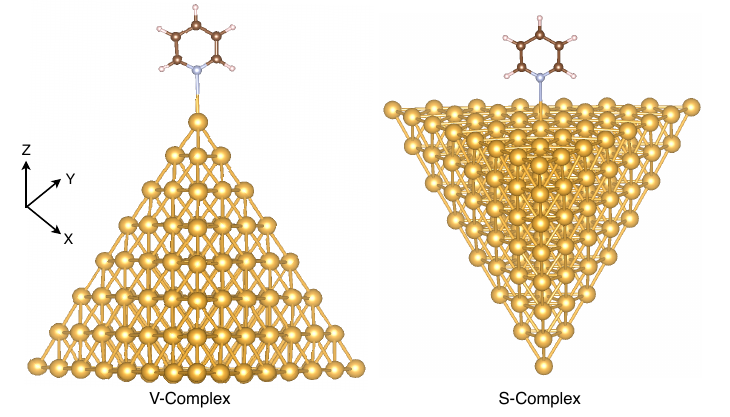}
    \caption{Optimized geometries of Au$_{120}$-pyridine complexes where pyridine is adsorbed on one of the tips (V-complex) and triangle surfaces (S-complex).}  
   \label{sys_fig}
\end{figure}

\begin{figure}[h]
    \centering
    \includegraphics[width = 0.78 \textwidth]{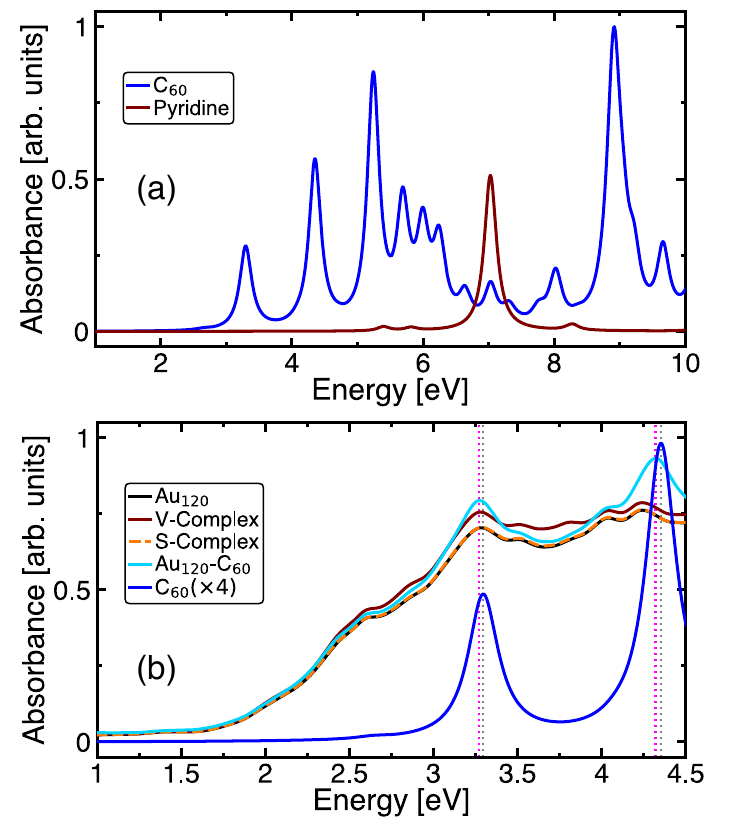}
    \caption{Absorption spectra of (a) isolated fullerene (C$_{60}$) and pyridine molecules, and (b) Au$_{120}$ complexes.  
     In (b) peak positions for the first two lowest energy peaks (gray dotted lines) of the isolated C$_{60}$ are compared with the peaks that appear for the Au$_{120}$-C$_{60}$ complex (magenta dotted lines) near 3.3 and 4.4 eV. 
    Absorption intensity for (a) and (b) are scaled differently.}  
   \label{spec_fig}
\end{figure}

\section{Results and Discussion} 
\subsection*{Raman Scattering}
\subsubsection*{Au$_{120}$-Pyridine Complexes:} 
We use a tetrahedral Au$_{120}$ nanoparticle for all of these studies, which is a similar structure (but larger) to what has been considered in earlier SERS studies.\cite{TDDFT-Schatz-2006,TDDFT-Schatz-2006} This allows two distinct binding sites for an adsorbing pyridine molecule, either on one of the tips (V-complex) or in the middle of the triangle faces (S-complex).
These two sites provide significantly different chemical environments for pyridine, affecting how the molecule interacts with the particle.  
It has been found that the S-complex better represents the experimental SERS spectra, however, the V-complex provides different SERS features due to a unique interaction between the particle and molecule.  \\

We optimize all geometries with a maximum force tolerance of 10$^{-5}$ a.u. for the movement of atoms. 
We attach pyridine on the tip (for V-complex) and surface (for S-complex) gold atoms of a previously optimized Au$_{120}$ and further optimize, freezing the atoms of the particle. 
Optimized geometries are shown in Figure \ref{sys_fig}. 
For both the geometries pyridine binds through the N atom and the Au-N bond aligns with the Z-axis. 
The distance between N and the nearest Au is 2.21 ${\rm \AA}$ for the V-complex and 2.32 ${\rm \AA}$ for the S-complex. 
A smaller distance between the particle and molecule was also observed in a Au$_{20}$-pyridine V-complex with TD-DFT calculations.\cite{TDDFT-Schatz-2006}  
We find a small amount of electron transfer from pyridine to the particle in the ground state structures, 0.05e and 0.03e for the V- and S-complexes, respectively. 
Furthermore, we note that an electronic temperature of 300K is used throughout the study.\\

\subsubsection*{Absorption Spectra:} 
The absorption spectrum of pyridine is shown in Figure \ref{spec_fig}(a). 
Excited states between 5-6 eV are weakly absorbing while a strong absorption peak appears at an energy of 7 eV. 
The strong absorption peak involves $\pi\rightarrow\pi^*$ transition and is associated with a large oscillator strength.
These spectral features are consistent with previous studies based on INDO/SCI and TD-DFT methods.\cite{INDO-Schatz-2017} 
The absorption spectrum of fullerene (Fig.\ref{spec_fig}(a)) is discussed in the next section. \\

Absorption spectra for Au$_{120}$ and Au$_{120}$-pyridine complexes are shown in Figure \ref{spec_fig}(b). 
The spectrum for the gold particle shows multiple peaks that are comparable to the spectrum of a tetrahedral Au$_{20}$ investigated using the TD-DFT method.\cite{TDDFT-Schatz-2006}  
It contains contributions from several intraband transitions at lower energies and interband transitions at relatively higher energies.  
However, the overall spectral intensity is increased with the particle size due to the large density of states available in Au$_{120}$. 
A weak LSPR peak appears at around 2.5 eV.
We note that although a gold particle involving 120 atoms is large compared to the previous study with Au$_{20}$, it is still too small sustain significant LSPR excitation. 
Nevertheless, excitation near 2.5 eV with dominant intraband transitions can generate results better resembling the LSPR characteristics that are usually observed experimentally for much larger particles.  
Note that the participation of $d$-electrons in photoexcitation causes decoherence of plasmon excitation, resulting in a broad plasmon peak that overlaps with inter-band transitions at higher energy. 
This makes it difficult to identify the LSPR peak for gold particles with a nearly spherical shape with less than about 300 atoms in the cluster, but for tetrahedral particles, the LSPR peak is apparent in our 120 atom results. 
The spectrum for the S-complex is very similar to the isolated gold particle spectrum, whereas for the V-complex the absorption intensity is slightly increased due to CT excitations as is consistent with previous TD-DFT results.\cite{TDDFT-Schatz-2006} \\

\subsubsection*{Enhancement of Raman Scattering:} 
We have used DFTB to compute the SERS enhancement for pyridine using low frequency (nominally static) calculations to determine the chemical enhancement factor, and with calculations at the plasmon energy of 2.5 eV for Au$_{120}$-pyridine complexes to determine the electromagnetic enhancement factor.
We find that resonance CT excitations for the Au$_{120}$-pyridine complexes involve small transition strengths that overlap with the dominant $d$-band excitations and therefore can have negligible contributions to SERS. 
Thus in the following, we only consider enhancements due to static chemical and electromagnetic effects. \\

\begin{figure}[h]
    \centering
    \includegraphics[width = 0.8 \textwidth]{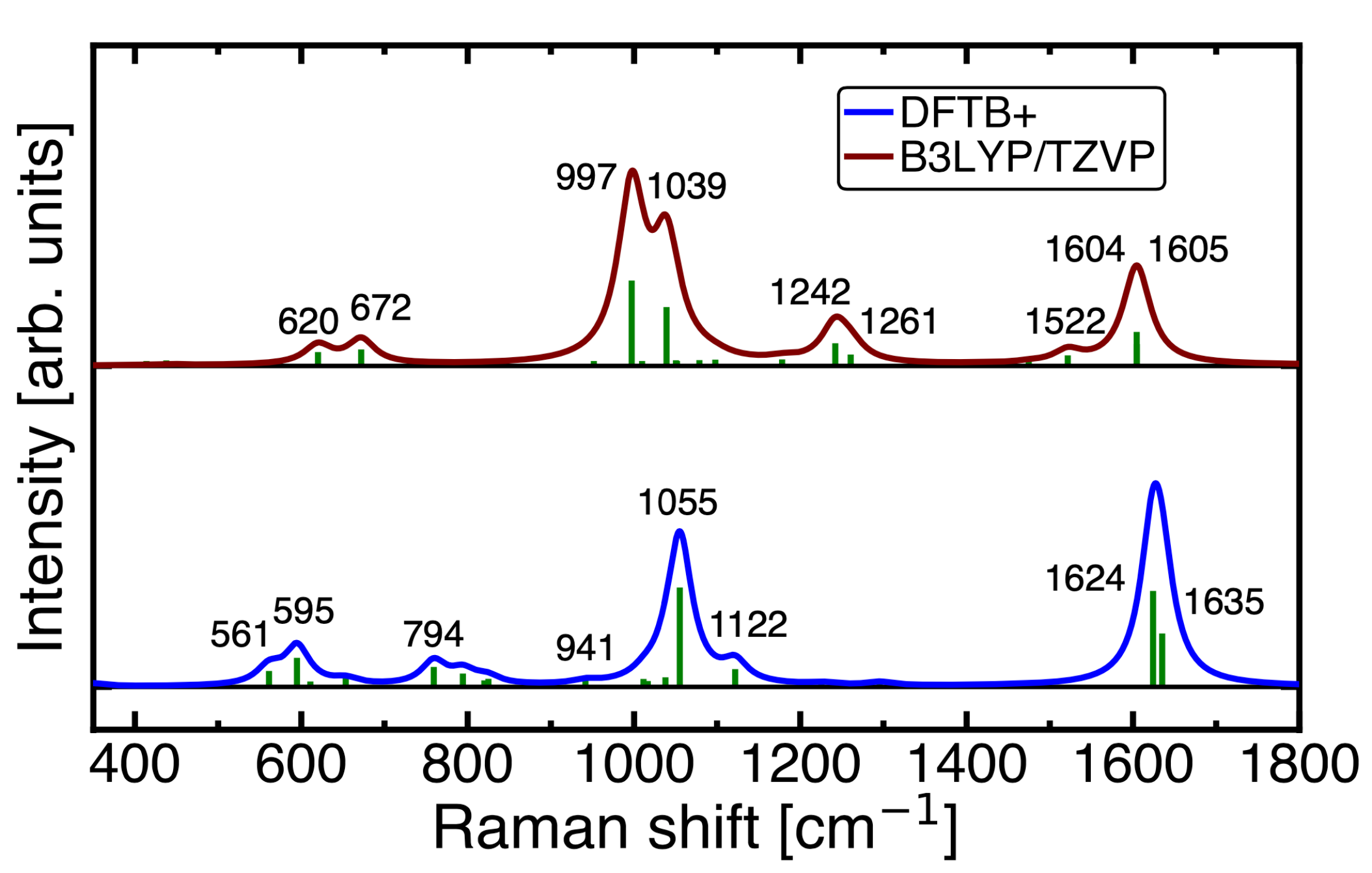}
    \caption{Raman spectra of pyridine using B3LYP/TZVP and TD-DFTB methods.}  
   \label{raman_b3lyp_fig}
\end{figure}

Normal Raman scattering spectra (obtained through computation of the static polarizability derivatives) for gas phase pyridine are shown in Figure \ref{raman_b3lyp_fig}, with the TD-DFTB result compared with the spectrum obtained from an analogous B3LYP/TZVP DFT calculations.
Here we have scaled the normal mode frequencies from DFTB by a factor of 0.9 to correct for a systematic overestimation of the vibrational frequency obtained from this method.  
Overall the spectra agree reasonably well both for the mode frequencies and relative peak intensities. 
The Raman active modes from the TD-DFT calculation include a pair of ring deformation modes at 620 and 672 cm$^{-1}$, a pair of ring breathing modes at 997 and 1039 cm$^{-1}$, a symmetric C-H wag mode at 1242 cm$^{-1}$, and a pair of ring stretching modes at 1604 and 1605 cm$^{-1}$.
The two noticeable discrepancies between spectra are (i) unphysical modes that appear in the TD-DFTB results near 794 cm$^{-1}$ and (ii) for the lower energy ring breathing mode (that appears at 941 cm$^{-1}$ in the TD-DFTB result) the intensity is significantly suppressed compared to the others.  
As will be discussed later the unphysical modes near 794 cm$^{-1}$ are not enhanced in SERS, however, the lower energy ring breathing mode can be significantly enhanced even though it is vanishingly small for the isolated gas phase molecule (TD-DFTB result). \\ 

\begin{figure}[h]
    \centering
    \includegraphics[width = 0.8 \textwidth]{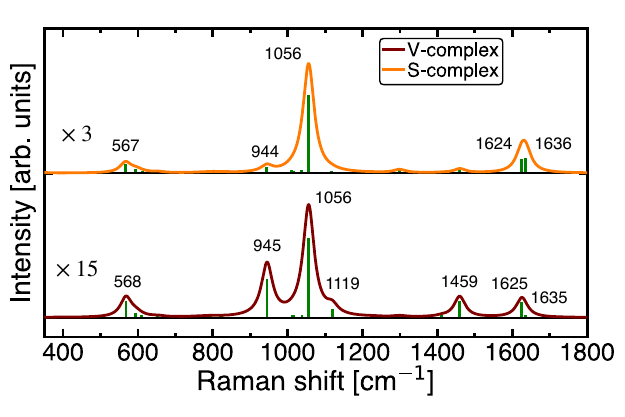}
    \caption{Surface enhanced Raman scattering for Au$_{120}$-pyridine complexes using the static polarizability of the molecule.
    Throughout the scaling factors in the inset represent enhancements compared to the normal Raman intensities of an isolated pyridine.}  
   \label{raman_static_fig}
\end{figure}

\begin{figure}[h]
    \centering
    \includegraphics[width = 0.8 \textwidth]{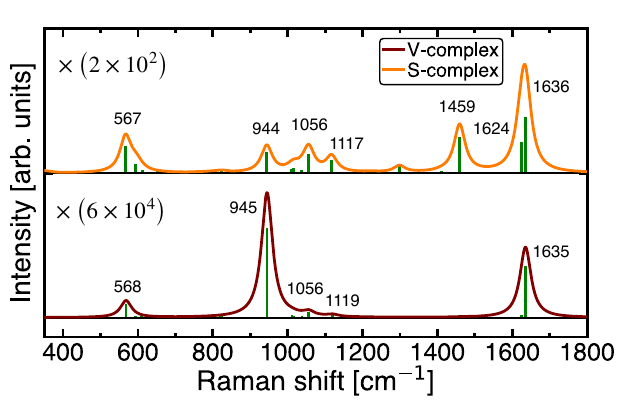}
    \caption{Surface enhanced Raman scattering of Au$_{120}$-pyridine complexes for an incident laser energy of 2.5 eV.}  
   \label{raman_res_fig}
\end{figure}

The normal Raman scattering spectra of pyridine adsorbed on the surface of Au$_{120}$ are shown in Figure \ref{raman_static_fig}.
These results show that the chemical enhancements are around a factor of 3 for the S-complex and 15 for the V-complex. 
These enhancement factors are comparable with the reported values (a factor 3 for the S-complex and 20 for the V-complex) for a Au$_{20}$-pyridine complex based on earlier TD-DFT calculations.\cite{TDDFT-Schatz-2006} 
The larger enhancement and sensitivity for the V-complex is understandable as the chemical interaction between particle and molecule is larger for that complex compared to the other given that the pyridine interacts with a coordinately less saturated gold atom.  \\

Interestingly the lower energy ring breathing mode (944 cm$^{-1}$) that has a small intensity for the isolated molecule (Fig.\ref{raman_b3lyp_fig}) is now enhanced by a factor of 15 for the V-complex, while the higher energy ring breathing mode (1056 cm$^{-1}$) shows a weak enhancement.
Between these two ring breathing modes the higher energy mode is enhanced more than that for the lower energy mode for the S-complex. 
Again, this trend in enhancements is consistent with the TD-DFT calculation as observed for Au$_{20}$-pyridine complexes.\cite{TDDFT-Schatz-2006}.
Further, the unphysical modes that appear in the isolated molecule near 794 cm$^{-1}$ have not been enhanced for either complex. 
Another mode with much higher enhancements is near 1459 cm$^{-1}$. This peak was also seen in the TD-DFT work, and it appears in experimental SERS measurements on gold but not silver.\cite{TDDFT-Schatz-2006}\\

The mode frequencies show blue-shifts for both complexes as observed in the TD-DFT calculations, however, the shifts are relatively smaller compared to the TD-DFT results.\cite{TDDFT-Schatz-2006}
The V-complex shows higher blue shifts than the S-complex for most of the Raman active modes, as also observed in the TD-DFT calculations.
The larger blue shifts in the V-complex can be associated with stronger bonding interaction in the V-complex as already mentioned. \\

The SERS scattering spectra of gold-pyridine for plasmon excitation at an energy of 2.5 eV are shown in Figure \ref{raman_res_fig} for both complexes.
The resonant excitation of metal complexes leads to enhancements due to electromagnetic effects along with chemical enhancements. 
The overall enhancement factors are on the order of $10^{1}-10^2$ and $10^3-10^4$ for the S- and V-complexes, respectively.
Similar enhancement factors were also observed for Au$_{20}$-pyridine complexes with the TD-DFT method.\cite{TDDFT-Schatz-2006} For the V-complex, the 941 and 1635 cm$^{-1}$ modes are enhanced significantly,  by a factor of $8\times10^{4}$ and  $4\times10^{4}$, respectively, while for the S-complex multiple modes are enhanced, with the maximum enhancement for 1636 cm$^{-1}$ mode. 
Further, a new Raman active mode appears in the S-complex at 1459 cm$^{-1}$.  
Interestingly this mode was strongly enhanced in the static Raman spectra and in the TD-DFT results noted above. \\

\begin{figure}[h]
    \centering
    \includegraphics[width = 0.8 \textwidth]{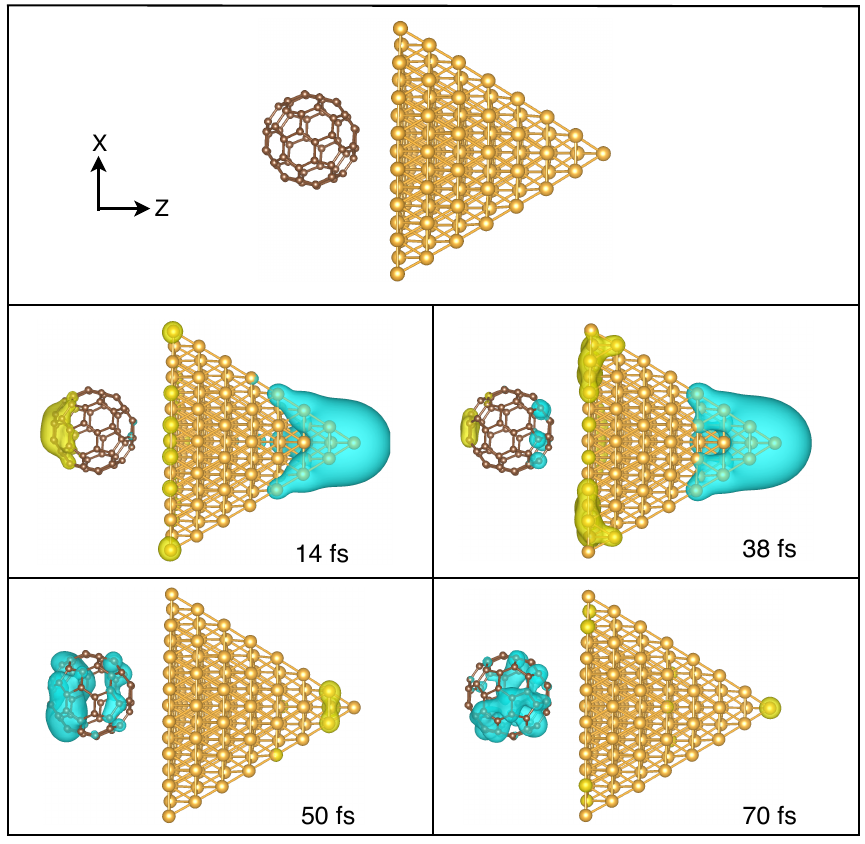}
    \caption{ (Top) Optimized geometry of a Au$_{120}$-C$_{60}$ complex, and (Bottom) snapshots of charge density difference with respect to the initial charge at $t=0$. 
    The isosurface cutoff is set to $\pm$0.05e.
    Yellow color represents positive and cyan color represents negative amplitudes.}  
   \label{sys2_fig}
\end{figure}

\begin{figure}[h]
    \centering
    \includegraphics[width = 0.8 \textwidth]{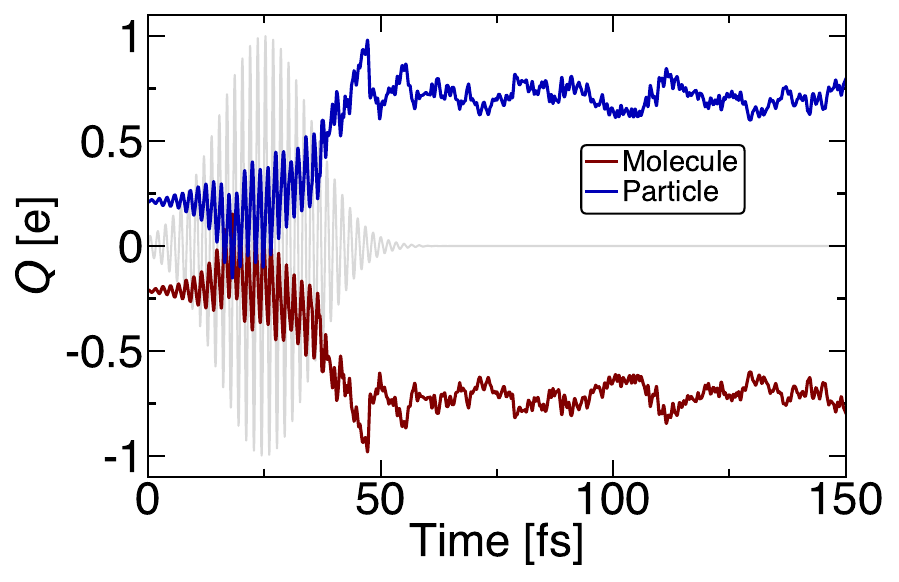}
    \caption{Time evolution of charges for particle and molecule for interaction with a Gaussian laser pulse shown by a light gray line in the background.}  
   \label{charge_fig}
\end{figure}

\subsection*{Charge Transfer}
\subsubsection*{Au$_{120}$C$_{60}$ Complex:}
To generate a suitable geometry to study plasmon-driven charge transfer, we place a fullerene (C$_{60}$) molecule on one of the triangular faces (as in the S-complex) of Au$_{120}$ (previously optimized) and then the geometry is optimized freezing atoms of the particle. 
The optimized geometry of Au$_{120}$C$_{60}$ is shown in Figure \ref{sys2_fig}.
The minimum distance between C and Au atoms is around 3.5 ${\rm \AA}$. 
We observe a small electron transfer (around 0.2e) from the particle to the molecule for the optimized geometry in the ground state. \\

\subsubsection*{Absorption Spectra:}
Absorption spectra for C$_{60}$ and Au$_{120}$C$_{60}$ are displayed in Figures \ref{spec_fig} (a) and (b), respectively.
The molecular spectrum exhibits multiple peaks, with the first three peaks at 3.3, 4.35, and 5.2 eV.
These calculated peaks well represent the experimental peaks found at 3.8, 4.9, and 6 eV, respectively for the gas phase C$_{60}$ molecule.\cite{C60-Shobatake-1996}  
The spectrum of the Au$_{120}$C$_{60}$ complex is dominated by the gold particle with two major peaks near 3.27 and 4.3 eV.
These two peaks are slightly red-shifted compared to the first two lowest energy peaks of the isolated molecule.
Previous studies demonstrated a shift in LSPR peaks due to the interaction between plasmon and molecular resonances.\cite{Plasmon-Molecule-VanDuyne-2006,Plasmon-Molecule-VanDuyne-2007}
Moreover, a weak plasmon peak near 2.5 eV is also observed as in the other complexes in the figure. \\ 

\subsubsection*{Laser-Induced Charge Transfer Dynamics:}
A linearly polarized (polarized along the Z-axis) laser pulse of the form given in Eq.\ref{pulse_eq} is used to drive the system.
Both molecule and particle couple with the external field. 
We use a driving field amplitude $E_0$, frequency $\omega_l$, and pulse duration $T_l$ of 0.7 V/${\rm \AA}$, 2.5 eV and 14 fs, respectively.
We note that for this value of the pulse duration, the field amplitude decays to zero at about 50 fs with a peak position at 25 fs. 
Figure \ref{charge_fig} displays the time evolution of charges for the molecule and particle separately. 
We note that the total charge is preserved (here charge neutral complex) at all times, however, charge transfer can occur across the interface between the molecule and particle during the dynamics. 
As seen in Fig.\ref{charge_fig}, the initial charge (before interacting with a laser pulse) of the molecule is slightly negative, reflecting ground state charge transfer to the molecule. 
The charges of the two subsystems oscillate with the laser frequency during interaction with the laser pulse. 
There is an effective transfer of electrons from the molecule to the particle during the early time of the dynamics (Figure \ref{sys2_fig}), and the charge of the particle drops to about +0.15e at around 18 fs. 
After that the electron transfer changes direction and an effective transfer of electrons from the particle to the molecule is observed.
At long times (after 50 fs), the charge on the molecule oscillates around a value of -0.75e.
Figure \ref{sys2_fig} displays the negative charge on the molecule at long times (after 50 fs).
The transfer of electrons and the formation of negative molecular ions for fullerene-based molecules have been experimentally demonstrated using SERS measurements, where the fullerene pentagonal pinch mode was found to change frequencies from that for the neutral fullerene to that for the fullerene anion after excitation of the gold/fullerene complex at 532 nm.\cite{SERS-PCBM-Schatz-2019} 
We note that since the fullerene is still in contact with the gold nanoparticle at the end of the calculation the amount of charge residing on the molecule is not necessarily a full electron.

\section{Conclusions}
In this study, we have presented two interrelated projects that test the ability of TD-DFTB to describe the influence of plasmon excitation on two physical phenomena: surface-enhanced Raman scattering (SERS) and plasmon-enhanced charge transfer. 
In the first, we calculated SERS spectra for Au$_{120}$-pyridine complexes, where we found enhancements due to both chemical and plasmon excitation effects that are consistent with earlier TD-DFT studies.\cite{TDDFT-Schatz-2006}
The enhancement factors for chemical effect are around 3-15 depending on the site of the adsorption.
For the electromagnetic effect, the enhancement is on the order of $10^3-10^4$ for the V-complex (for adsorption on one of the tips) whereas $10^1-10^2$ for the S-complex (for adsorption on one of the triangle faces). 
In the second project, we studied plasmon-driven charge transfer across particle-molecule junctions after laser excitation of the particle near the plasmon energy at 2.5 eV.
We found substantial electron transfer from particle to molecule for a Au$_{120}$-C$_{60}$ complex leading to the formation of a negative molecular ion, C$_{60}^{-}$ on the particle.
This observation is consistent with a previous experimental result that shows evidence of electron transfer from particle to molecule and the formation of negative molecular ions for a fullerene derivative.\cite{SERS-PCBM-Schatz-2019}
Noteworthy in this result is that we did not see other photoprocesses in these simulations of laser pulses with intensities that are close to threshold, including C-C bond breakage, fragmentation of the particle, or desorption of the fullerene from the particle.
The ability of the DFTB method to simulate SERS and charge transfer problems opens several new possibilities in the plasmonics field.  

\section*{Acknowledgements}
This research was supported by the Office of Basic Energy Science, Department of Energy, through grant DE-SC0004752.

\printbibliography[title={References}]
\end{document}